\documentclass[prl,showpacs,amsmath,preprint]{revtex4}  
\usepackage{amssymb}
\usepackage{amsmath}

\begin{document}

\title{Diffusion of optical pulses in dispersion-shifted randomly birefringent optical fibers}

\author{Pavel M. Lushnikov$^{1,2}$
}

\address{$^1$ Theoretical Division, Los Alamos National Laboratory,
  MS-B213, Los Alamos, New Mexico, 87545
  \\
  $^2$ Landau Institute for Theoretical Physics, Kosygin St., 2,
  Moscow, 119334, Russia
  }
\email{lushnikov@cnls.lanl.gov}


\begin{abstract}
An effect of polarization-mode dispersion, nonlinearity and
random variation of dispersion along an optical fiber on a pulse
propagation in a randomly birefringent dispersion-shifted optical
fiber with zero average dispersion is studied. An averaged pulse
width is shown analytically to diffuse with propagation distance
 for arbitrary strong pulse amplitude. It is
found that optical fiber nonlinearity can not change
qualitatively a diffusion of pulse width but can only modify a
diffusion law which means that a root mean square pulse width
grows at least as a linear function of the propagation distance.
\end{abstract}



\maketitle


Polarization-mode dispersion (PMD), which is a pulse broadening
caused by random variation of optical fiber birefringence, has
recently become a major drawback in the development of new
high-bit-rate optical communication systems
\cite{Poole1988,89Mol,89Men,Poole1991,Gisin1992,96Men,YangKathMenyuk2001}.
Another effect, which limits bit-rate capacity, is pulse
broadening caused by group-velocity dispersion (GVD). Use of a
dispersion-shifted fiber with zero average GVD can reduce this
effect, however, in such fibers GVD inevitably fluctuates around
zero along the propagation direction \cite{96MMN,00MMGNMGV} and
hence pulse broadening still occurs
\cite{ChertkovGabitovMoeser2001,ChertkovGabitovLushMT2002}.
Nonlinearity in optical fibers results in the coupling of both PMD
and GVD effects, so in general they can not be studied separately
in contrast to linear case. Linear PMD was first studied in Refs.
\cite{Poole1988,Poole1991,Gisin1992} while nonlinear PMD was
addressed in numerical experiments \cite{89Mol} and analytical
studies based on a perturbation expansions around soliton
solutions of deterministic equations
\cite{89Men,LakobaKaup1997,Matsumoto1997,chertkovgabitovjetplett2001b,Chertkov2004}.
An effect of random variation of GVD was studied in Refs.
\cite{ChertkovGabitovMoeser2001,ChertkovGabitovLushMT2002,Abdullaev2000a,Abdullaev2000b,chertkovgabitovjetplett2001a}.
 Here an exact analytical (nonperturbartive) theory is developed
for the case of fiber with random birefringence and random GVD
with zero mean and arbitrarily strong nonlinearity (arbitrary
pulse amplitude). No assumption like closeness to any type of
soliton solution is necessary for the results of this Article to
be valid. The main result is that a statistical average (over
random variation of fiber parameters) of root mean square pulse
width $T_{RMS}$ grows with distance at least as a linear function
of propagation distance. This means that random diffusion of
optical pulse width can not be prevented by an arbitrarily  strong
nonlinearity. It is shown that random diffusion fundamentally
limits the bit-rate capacity of an optical fiber.

Neglecting second-order GVD (dispersion slope) effects,
stimulated Raman scattering and Brillouin scattering, the
propagation of optical pulses in birefringent optical fibers is
described by the two-component vector nonlinear Schr\"odinger
equation
(VNLS)\cite{zakharovberkhoer1970,89Men,97Agrawal,chertkovgabitovjetplett2001b}
\begin{eqnarray}
i \partial _z\eta_\alpha +\sum_{\beta=1}^2{
\triangle}_{\alpha\beta}(z)\eta_\beta+i\sum_{\beta=1}^2\tilde
m_{\alpha\beta}(z)
\partial _t\eta_\beta+d(z)\partial^2 _t\eta_\alpha \nonumber \\   + \sigma(z)
\tilde N_\alpha(\eta) = i G(z)\eta_\alpha, \label{vnls0}
\end{eqnarray}
where $z$ is the propagation distance along an optical fiber,
$\eta_1$ and $\eta_2$ correspond to the complex amplitudes of  two
orthogonal linear polarizations, $t\equiv \tau-z/c_l$ is the
retarded time and $\tau$ is the physical time, $c_l$ is the speed
of light, and $d(z)$ is the dispersion, which is related to
first-order GVD $\beta_2$ as $d(z)=-\frac{1}{2}\beta_2(z)$. The
right hand side (rhs) of Eq. $(\ref{vnls0})$ describes linear
losses and amplifiers, $G(z)\equiv\Big(-\gamma
+[\exp{(z_a\gamma)}-1]\Sigma^N_{k=1}\delta(z-z_k)\Big ),$
$\sigma=(2\pi n_2)/(\lambda_0 A_{eff})$ is the nonlinear
coefficient, $n_2$ is the nonlinear refractive index,
$\lambda_0=1.55\mu m$ is the carrier wavelength, $A_{eff}$ is the
effective fiber area, $z_k=kz_a\ (k=1,\ldots, N)$ are the
amplifier locations, $z_a$ is the amplifier spacing, and $\gamma$
is the loss coefficient. Distributed amplification can be also
included by adding $z$-dependence into $\gamma$. Properties of
fiber can be different along optical line, e.g. $A_{eff}$ could
be different if line consists of several pieces of fiber with
different cross section, and, respectively, coefficient $\sigma$
generally depends on $z$. In a similar way, all parameters of
fiber, like $d(z)$ also depend on $z$.

The self-conjugated matrices $\hat \triangle(z)$ and $\hat
{\tilde m}(z)$ describe, respectively, the differences in wave
vectors and
 the anisotropy of the group velocities of the two modes corresponding to
the two different polarizations. Both matrces $\hat \triangle$
and $ \hat{\tilde m}$ are made traceless. The trace of the matrix
$\hat \triangle$ is excluded by a phase transformation $\tilde
\eta\to \eta\exp(i\phi_0z)$. The trace of the matrix $\hat{\tilde
m}$ is zero because Eq. $(\ref{vnls1})$ is written in a frame
moving with average group velocity (note that group velocity is
generally $z$-dependent).
  It is
assumed in Eq. $(\ref{vnls1})$ that the dispersion $d(z)$ and
nonlinearity are isotropic because their anisotropy is usually
negligible in optical fibers. Vector $\tilde{\bf N}=(\tilde
N_1,\tilde N_2)^{T}$, which represents the contribution of Kerr
nonlinearity, is given by:
\begin{eqnarray} \label{Ndef0}
\tilde N_1(\Psi)=\Big [(|\Psi_1|^2+\frac{2}{3}|\Psi_2|^2)\Psi_1+\frac{1}{3}\Psi_2^2\Psi_1^* \Big], \nonumber \\
\tilde N_2(\Psi)=\Big
[(\frac{2}{3}|\Psi_1|^2+|\Psi_2|^2)\Psi_2+\frac{1}{3}\Psi_1^2\Psi_2^*
\Big]
\end{eqnarray}
(see \cite{zakharovberkhoer1970,89Men}).

 The change of variables $
\xi=\eta e^{ -\int^z_0 G(z')dz'}$ (see e.g. Refs.
\cite{gabtur1,lush2002a}) removes rhs of Eq. $(\ref{vnls0})$ and
gives:
\begin{eqnarray} \label{vnls1}
i \partial _z \xi_\alpha +\sum_{\beta=1}^2{
\triangle}_{\alpha\beta}(z)\xi_\beta+i\sum_{\beta=1}^2\tilde
m_{\alpha\beta}(z)
\partial _t\xi_\beta\nonumber \\+d(z)\partial^2 _t\xi_\alpha
 +c(z)\tilde N_\alpha(\xi)=0,
\end{eqnarray}
where $c(z)\equiv \sigma(z)\exp{\Big(2\int^z_0G(z')dz'\Big)}$.
Thus all linear fiber losses and amplifications are included into
coefficient $c(z).$

The isotropic case, which corresponds to zero matrices $\hat
\triangle=\hat{\tilde m}=\hat 0$, allows a solution of Eq.
$(\ref{vnls1})$ with constant polarization, e.g. $\xi_1\ne0,
\xi_2=0$. Components of matrices $\hat \triangle$ and $\hat{\tilde
m}$  fluctuate strongly as  functions of distance $z$.
Fluctuations correspond to violation of circular symmetry of the
fiber. The matrices $\hat \triangle$ and $\hat {\tilde m}$  change
in optical fibers  with time on a scale of few hours because of
environmental fluctuations, however, for typical optical pulse
duration (10ps), one can consider $\hat \triangle$ and
$\hat{\tilde m}$ as functions of $z$ only. It means that disorder
is frozen in the fiber. The matrix $\hat \triangle$ gives the
leading order contribution in Eq. $(\ref{vnls1})$ because a
typical beat length $z_{beat}$ (typical length at which a relative
phase shift between two polarizations caused by $\hat \triangle$
becomes $\sim \pi$)  is $\sim 20m$ in optical fibers
\cite{96Men,LakobaKaup1997}. Contribution of all other terms in
Eq. $(\ref{vnls1})$ are essential on a scale of the order of
$10km$ and larger. Thus, it is convenient to introduce a slow
variable $\Psi$ as
\begin{eqnarray} \label{psidef1}
\eta_\alpha=\sum_{\beta=1}^2T_{\alpha \beta}(z)\Psi_{\beta},
\end{eqnarray}
where the unitary matrix $\hat T(z)$ is given by the solution of
the matrix equation:
\begin{eqnarray} \label{Tdef1}
\frac{dT_{\alpha \beta}}{dz}=i\sum_{\delta=1}^2\triangle_{\alpha
\delta}(z)T_{\delta \beta}, \quad  T_{\alpha \beta}(0)={\hat I},
\end{eqnarray}
where ${\hat I}$ is the identity matrix.

The slow variable $\Psi$ changes on the scale  $\sim 10 km$
because all fast dependence on $\hat \triangle$ is included into
$\hat T$. Thus Eq. $(\ref{vnls1})$ can be averaged over distance
much larger than $z_{beat}$, but still much smaller than $10km$
\cite{89Men,96Men,chertkovgabitovjetplett2001b,Chertkov2004}. Here
we use the simplest possible form of an averaged equation
\cite{89Men,chertkovgabitovjetplett2001b,Chertkov2004}:
\begin{eqnarray} \label{vnls2}
i \partial _z \Psi_\alpha +i\sum_{\beta=1}^2m_{\alpha\beta}(z)
\partial _t\Psi_\beta+d(z)\partial^2 _t\Psi_\alpha
 + N_\alpha=0,
\end{eqnarray}
where $\hat m(z)=\hat T^{-1}\cdot \hat {\tilde m}(z)\cdot \hat T$
and the nonlinear terms are
\begin{eqnarray} \label{Ndef1}
 N_1=c(z)(|\Psi_1|^2+\frac{2}{3}|\Psi_2|^2)\Psi_1,\nonumber \\
 N_2=c(z)(\frac{2}{3}|\Psi_1|^2+|\Psi_2|^2)\Psi_2.
\end{eqnarray}
 The typical
scale $z_{corr}$ of  variation of $\hat m$ in optical fibers is
between 10m and 100m. Typical scales at which a pulse experiences
essential distortion are a dispersion length $z_{disp}\equiv
t_0^2/d_1$, $d_1$ is the typical amplitude of dispersion
variations, $t_0$ is the typical pulse width; a nonlinear length
$z_{nl}\equiv 1/(c(z)p^2)$, $p$ is the typical pulse amplitude;
and PMD length $z_{m}\equiv t_0/m_0$, $m_0$ is the typical
amplitude of variation of matrix $\hat m$ components. As a
typical example one can set $t_0\sim 10ps, \, d_1\sim 1ps^2/km, \,
p^2\sim 2mW, c(z)\sim 0.001 \, (km \, mW)^{-1},$ and $m_0\sim 1
ps/km$.  One gets $z_{disp}\sim 100km, \, z_{nl}\sim 500 km,$ and
 $z_{m}\sim 10km$. Thus, a minimal length of pulse distortion
$z_{pulse}$ is $\sim z_{m}\sim 10km$. the length $z_{pulse}$ is
much larger than $z_{corr}$ and, according to the central limit
theorem, $\hat m$ can be treated at scales of the order of
$z_{pulse}$ as random Gaussian processes with zero correlation
length and zero mean $\langle \hat m\rangle=\hat 0$ ($\langle
\ldots \rangle$ means ensemble averaging over statistics of $\hat
m(z)$ and $d(z)$).

The traceless matrix $\hat m$ can be represented in terms of Pauli matrices:
%
$ \hat m(z) =\sum_{j=1}^3m_j(z) \hat \sigma_j,
$
%
where the correlation functions of components of the real vector
${\bf m}$ are given by
\begin{eqnarray} \label{MDdef2}
\langle m_j(z_1)m_k(z_2)\rangle=M_j\delta_{jk}\delta(z_1-z_2).
\end{eqnarray}
The vector ${\bf M}$ does not depend on $z$ and ${\bf M}$ is
defined from the original problem with short but nonzero
correlation lengths as $M_j=\int \langle m_j(z) m_j(z')\rangle
dz'$.

Similarly, dispersion $d(z)$ in a dispersion-shifted fiber (where
average GVD is shifted to zero but random variations of GVD are
essential \cite{96MMN,00MMGNMGV,ChertkovGabitovLushMT2002}) can be
represented by random Gaussian process with zero correlation
length and zero mean:
\begin{eqnarray} \label{dcorr1}
\langle d(z_1)d(z_2)\rangle=D\delta(z_1-z_2), \quad \langle d(z)\rangle=0,
\end{eqnarray}
where $D=\int \langle d(z) d(z')\rangle dz'$. The quantities $\hat
m(z)$ and $d(z)$ are assumed to be the independent random
processes: $\langle m_j(z) d(z')\rangle=0$.

Below the Eqs. $(\ref{vnls2})-(\ref{dcorr1})$ are used to obtain
exact result on evolution of pulse width along $z$. First one can
conclude that as $\langle \hat m(z)\rangle=\hat 0$ and  $\langle
d(z)\rangle=0$, there is no preferred direction along $t$ and the
average pulse position is zero:
\begin{eqnarray} \label{Taver}
\langle t\rangle \equiv \langle\int
t(|\Psi_1|^2+|\Psi|^2)dt\rangle/P=0,
\end{eqnarray}
where
\begin{eqnarray} \label{Power1}
P= \int (|\Psi_1|^2+|\Psi_2|^2)dt
\end{eqnarray}
is the time-averaged optical power. The quantity $P$ is an
integral of motion of VNLS $(\ref{vnls2}):$ $\partial _z \,P_z=0.$

Consider now evolution along $z$ of
\begin{eqnarray} \label{Adef1}
A\equiv \int (t-\langle
t\rangle)^2(|\Psi_1|^2+|\Psi|^2)dt,\end{eqnarray}
which is related to the root mean square pulse width $T_{RMS}$ by
\begin{eqnarray} \label{TRMSdef}
T_{RMS}^2=A/P.
\end{eqnarray}

 Using Eqs.  $(\ref{vnls2})$ and $(\ref{Taver})$, integrating
by parts over $t$, and applying vanishing boundary conditions at
infinity one derives
\begin{equation}\label{Az}
A_z=d(z)B^{(d)}-\sum_{j=1}^3m_j(z) \,B^{(m_j)},
\end{equation}
where
\begin{eqnarray}\label{Bdef1}
B^{(d)}=\int 2 i t \sum_{\alpha=1}^2(\Psi_\alpha \partial_t
\Psi_\alpha^\ast-\Psi_\alpha^\ast\partial_t\Psi_\alpha)dt,\nonumber\\
B^{(m_1)}=-2\int t (\Psi_1^*\Psi_2+c.c.)dt,\nonumber\\
B^{(m_2)}=2i\int t (\Psi_1^*\Psi_2-c.c.)dt, \\
B^{(m_3)}=-2\int t (|\Psi_1|^2-|\Psi_2|^2)dt.\nonumber
\end{eqnarray}
It is essential that all expressions in rhs of $(\ref{Bdef1})$ do
not have explicit dependence on random variables $d$ and ${\bf
m}$, which allows one to differentiate them over $z$ and find,
using again Eq. $(\ref{vnls2})$ and integrating by parts, that
\begin{subequations}
\begin{eqnarray}
 B^{(d)}_z=8d(z)X-2c(z)Y+ O({\bf m}),\label{Bz1a} \\
B^{(m_1)}_z=-2m_1P+O(d,m_2,m_3,c(z)), \label{Bz1b} \\
B^{(m_2)}_z=-2m_2P+O(d,m_1,m_3,c(z)),\label{Bz1c} \\
B^{(m_3)}_z=-2m_3P+O(d,m_1,m_2),\label{Bz1d}
\end{eqnarray}
\end{subequations}
where
\begin{eqnarray} \label{XYdef}
X\equiv \int\Big ( |\partial_t\Psi_1|^2+|\partial_t\Psi_2|^2\Big
) dt, \nonumber \\
Y\equiv \int
(|\Psi_1|^4+|\Psi_2|^4+\frac{4}{3}|\Psi_1|^2|\Psi_2|^2)dt.
\end{eqnarray}
Notation $O(d,m_2,m_3,c(z))$ in Eq. $(\ref{Bz1b})$ means extra
terms which are linear in $d,m_2,m_3,c(z)$. And the same type of
notation is used for the similar terms in rhs of Eqs.
$(\ref{Bz1a}),(\ref{Bz1c}),(\ref{Bz1d})$. Explicit expressions
for these terms are not given here because they are bulky and, as
it is shown below, they vanish after statistical averaging. Note
that for ${\bf m}=0, \ \Psi_2=0$ and  $d=Const$ Eq.
$(\ref{Bz1a})$ coincides with a so-called virial theorem
\cite{zakh1972,lush1995,lush2000b}. However, direct application
of the virial theorem to Eq. $(\ref{vnls2})$ is not possible
because it would require determination of $A_{zz}$ by
differentiating Eq. $(\ref{Az})$ over $z$, which would result in
appearance of terms including derivatives of random variables
$d(z)$ and ${\bf m}$ over $z$.  Here these problems are avoided by
studying  $ B^{(d)}_z$ and $B^{(m_j)}_z$ instead of $A_{zz}$, but
as a result of such procedure, one obtains cumbersome expressions
$(\ref{Bz1a})-(\ref{Bz1d})$ compare with the compact expression
for the virial theorem in deterministic scalar case  (see Refs.
\cite{zakh1972,lush1995,lush2000b}).

Consider the statistical average of
\begin{eqnarray}\label{Azd}
A_z^{(d)}=d(z)\Big(B^{(d)}(0)+\int^z_0 B^{(d)}_{z'}(z')dz'\Big ),
\end{eqnarray}
and substitute $B^{(d)}_{z'}(z')$ in that equation by rhs of Eq.
$(\ref{Bz1a})$. Assume for a moment (to choose a correct limit
$z_{corr}\to 0$) that the correlation length $z_{corr}$ of all
random processes $d(z)$ and ${\bf m}(z)$ is small but finite and
that all correlation functions decay at least exponentially or
faster with distance.  Now choose a distance $z_0$ in such a way
that $z_{corr}\ll z_0 \ll z_{pulse}$. The statistical average of
Eq. $(\ref{Azd})$, using Eq. $(\ref{Bz1a})$, gives
\begin{eqnarray}\label{Azdav}
\langle A_z^{(d)}\rangle=\langle d(z)\int^z_{z_0}\Big (8d(z')X(z')-2c(z')Y(z')\Big )dz'\rangle,
\end{eqnarray}
where all corrections to that equation are exponentially small
and vanish in the limit $z_{corr}\to 0$ due to casuality
constrain. Casuality constrain  implies here that $\Psi(z)$ does
not depend on $d(\tilde z)$ and  ${\bf m}(\tilde z)$ if $z<\tilde
z$.

The condition $z_0 \ll z_{pulse}$ allows one to write solution of
 VNLS $(\ref{vnls2})$ as $ \Psi(z')=\Psi(z_0)+O(z'-z_0)$ for
$z_0\le z'\le z. $ Taking the limit $z_{corr}\to 0$, for which
$z_0\to z$, and using Eq. $(\ref{Azdav})$ one obtains
\begin{eqnarray}\label{Azdavtot}
\langle A_z^{(d)}\rangle=4D\langle X\rangle.
\end{eqnarray}

To average Eq. $(\ref{Az})$ one rewrites it  in the equivalent
form
\begin{eqnarray}\label{Az2}
A_z=d(z)\Big(B^{(d)}(0)+\int^z_0 B^{(d)}_{z'}(z')dz'\Big )\nonumber\\
-\sum_{j=1}^3m_j(z)\Big(B^{(m_j)}(0)+\int^z_0
B^{(m_j)}_{z'}(z')dz'\Big ),
\end{eqnarray}
and substitute expressions for
$B^{(d)}_z,B^{(m_1)}_z,B^{(m_2)}_z,B^{(m_3)}_z$ from rhs of system
$(\ref{Bz1a})-(\ref{Bz1d})$ into Eq. $(\ref{Az2})$. As a result,
one finds, by averaging over distribution of ${\bf m}$ and $d,$
that
\begin{equation}\label{Azav1}
\langle A_z\rangle=P\sum_{j=1}^3 M_j+ 4D\langle X\rangle.
\end{equation}
Here $\langle P\rangle =P$ because $P$ is an integral o motion.
All terms in rhs of Eq. $(\ref{Azav1})$ are positive definite and
hence $\langle A\rangle$ grows with $z$.

Using Eqs. $(\ref{TRMSdef})$ and $(\ref{Azav1})$ one obtains an
expression for the statistical average of the root mean square
pulse width
\begin{equation}\label{Tzav1}
\langle T_{RMS}(z)^2\rangle=T_{RMS}(0)^2+z\sum_{j=1}^3 M_j+
\frac{4D}{P}\int_0^z\langle X(z')\rangle dz'.
\end{equation}

Eq. $(\ref{Tzav1})$ is exact for system
$(\ref{vnls2})-(\ref{dcorr1})$ because  Eq.  $(\ref{Tzav1})$ is
derived for arbitrary strong nonlinearity. Thus, this is
essentially nonperturbative result. Remarkably, Eq.
$(\ref{Tzav1})$ does not explicitly depend on the nonlinear
coefficient $c(z)$. This fact is very peculiar propertiy of VNLS
$(\ref{vnls2})$  and is related both to the generalization of the
virial theorem of scalar nonlinear Schr\"odinger equation
\cite{zakh1972,lush1995,lush2000b} and  to the zero correlation
length limit of fiber parameters $(\ref{MDdef2})$ and $
(\ref{dcorr1})$.
 In the linear case, which corresponds to $c(z)=0$, $X$ does not depend on $z$ and the
 growth of $\langle T_{RMS}(z)^2\rangle$
  with $z$ is
linear (i.e. the pulse width experiences diffusive growth caused
by random variations of GVD and PMD matrix $\hat m(z)$). For
nonzero $c(z)$, the nonlinearity is only responsible for
nontrivial dependence of $\langle X\rangle$ on $z$, and
therefore, for modification of the (still) diffusive law. The term
$P\sum_{j=1}^3 M_j$ is constant for arbitrarily strong
nonlinearity which means that growth of the pulse width can not
be slower than linear in distance $z$. This gives a fundamental
limit to the bit-rate of information transmission in optical fiber
systems.

Estimating $X$, $P$, $|{\bf M}|$ and $D$ by $p^2/t_0$, $p^2 t_0$,
$z_{corr} m_0^2$ and $z_{corr}d_1^2$, respectively, one derives
from Eq. $(\ref{Tzav1})$ that
$\beta\equiv \left(\langle T_{RMS}(z)^2\rangle-T_{RMS}(0)^2\right)/T_{RMS}(0)^2\nonumber \\
\sim zz_{corr}\Big (m_0^2/t_0^2+ d_{1}^2 /t_0^4\Big ). $ Then the
minimal requirement for small information loss, $\beta
\stackrel{<}{\sim} 1$, results in a limitation for the pulse
width: $t_0 \stackrel{>}{\sim} 10 ps$, for $z\sim 10^3 km$, and
the typical values $z_{corr}\sim 100m$, $m_0\sim 1ps/km$,  $d_{1}
\sim 1ps^2/km$. It suggests that construction of high-bit-rate
lines based on the dispersion-shifted fiber requires essential
improvement of fiber production technology, and/or implementation
of both PMD compensation \cite{Andrekson2001,Andrekson2002} and
the pinning method \cite{ChertkovGabitovMoeser2001}, to reduce
both PMD and random GVD variation effects.

Note that the extension of the results of this Article to the case
of both nonzero average dispersion (dispersion-shifted fiber with
nonzero average dispersion) and dispersion-managed systems is an
open problem. E.g. nonzero $\langle d\rangle =d_0(z)$ would
result in appearance of new terms $8d_0(z)\int^{z}_{z_0}\big
[d_0(z')\langle X(z')\rangle-2c(z')\langle Y(z')\rangle \big ]
dz'$ in rhs of Eq. $(\ref{Azav1})$ which generally are not
sign-definite. In the case of dispersion management these terms
oscillate fast with distance $z$ so their contribution to Eq.
$(\ref{Azav1})$ can be small. Additional research is necessary for
the case of nonzero average dispersion which is outside the scope
of this Article.

In conclusion, the exact analytical expression $(\ref{Tzav1})$ for
random diffusion of the averaged optical pulse width is derived.
Eq. $(\ref{Tzav1})$ introduces a fundamental limit on the minimal
pulse width for which information transmission is possible in
nonlinear dispersion-shifted optical fiber systems with zero
average dispersion.


The author thanks M. Chertkov, I.R. Gabitov and E.V. Podivilov for helpful discussions.

Support was provided by the Department of Energy, under contract W-7405-ENG-36.

E-mail address: lushnikov@cnls.lanl.gov.

\newpage





\end{document}